\documentclass [11pt,a4paper] {article}

\usepackage[cp1252]{inputenc}

\usepackage{amssymb}
\usepackage{amsmath}
\usepackage{amsfonts,amssymb}

\usepackage[dvips]{graphicx}

\usepackage{bbm}

\DeclareMathAlphabet{\mathpzc}{OT1}{pzc}{m}{it}

\begin{document}

\title{On the ongoing experiments looking for higher-order interference: What are they really testing?}

\author{Arkady Bolotin\footnote{$Email: arkadyv@bgu.ac.il$\vspace{5pt}} \\ \textit{Ben-Gurion University of the Negev, Beersheba (Israel)}}

\maketitle

\begin{abstract}\noindent The existence of higher than pairwise quantum interference in the set-up, in which there are more than two slits, is currently under experimental investigation. However, it is still unclear what the confirmation of existence of such interference would mean for quantum theory – whether that usual quantum mechanics is merely a limiting case of some more general theory or whether that some assumption of quantum theory taken as a fundamental one does not actually hold true. The present paper tries to understand why quantum theory is limited only to a certain kind of interference.\\

\end{abstract}

\section{Introduction}\label{Introduction}

\noindent What is the reason for quantum mechanics to be limited only to second-order interference? Indeed, in principle, one can consider a theory general enough to accommodate not only pairwise interference, but also higher types involving three or more alternatives (using, for example, a widely-studied framework known as the generalized probabilistic theory, GPT, \cite{Hardy, Barrett}). But if quantum mechanics is a special (i.e., limiting) case of the GPT framework, why doesn’t it exhibit higher-order interference? Naturally, one may assume that there is some mechanism by which the magnitude of higher-order interference is suppressed in quantum mechanics, but what is the origin of this mechanism?\\

\noindent Recently, several high-precision experiments, which measure the interference patterns produced by three, four and five slits and all the possible combinations of those slits being open or closed, have been completed searching for higher order interference \cite{Sinha, Sollner, Park, Kauten, Rengaraj}. These experiments were implemented in optics as well as via nuclear magnetic resonance in molecules, and they all but one have delivered results, which are in accordance with the presence of second-order interference only.\\

\noindent But then again, second-order interference is a consequence of the Born rule, which – as it is considered by many (see \cite{Deutsch, Assis, Zurek}, to name but a few) – is not independent but emergent from the rest of quantum formalism. In this way, experimentally confirmed existence of third- or higher-order interference would be evidence of the violation of certain principle (or principles) of that formalism.\\

\noindent The natural question that can arise then is, what particular (if any) fundamental assumption of quantum formalism does higher-order interference break?\\

\noindent Seeing as a GPT-based quantum theory that can supposedly shed light on possible pathological features of higher-order interference has not been realized (at least, yet) \cite{Lee}, to answer to that question a different approach is proposed in the present paper.\\

\section{Measuring quantum interference}\label{Measuring}

\noindent Specifically, to investigate into the problem of the absence of higher-order interference in quantum mechanics, the intuitionistic account of quantum interference is proposed, in which propositions concerning effects – such as ones which correspond to placing detectors directly behind slits in $N$-slit experiments or employing `telescopes' (like ones mentioned in Wheeler's original thought experiment \cite{Wheeler}) tightly aimed at the slits from some distance – obey constructive (intuitionistic) logic.\\

\subsection{A two-slit experiment}\label{Two-slit}

\noindent For the sake of clarity and simplicity, let us start with a double-slit experiment. Consider the following proposition:\smallskip

\begin{equation} \label{1} 
        \left(\{A,B\}\right)
        \equiv
        \big(
           \left(\{A\}\right) \oplus \left(\{B\}\right)
        \big)
   \;\;\;\;  ,
\end{equation}
\smallskip

\noindent where enclosed in parentheses expressions $\left(\cdot\right)$, which are capable of being `true' $\top$ or `false' $\bot$, denote propositions, the symbol $\oplus$ stands for the logical operation of \textit{exclusive disjunction} \cite{Germundsson} that outputs the logical value $\top$ only when its inputs $\left(\{A\}\right)$ and $\left(\{B\}\right)$ – the propositions of elementary events $\{A\}$ and $\{B\}$, i.e., subsets of the two-event space $\{A,B\}$$\,$\footnote{\label{f1}An event space is also referred to as sample space or possibility space.\vspace{5pt}} – differ, namely, $\left(\{A\}\right) \neq \left(\{B\}\right)$.\\

\noindent The detectors behind the slits $A$ and $B$ (or the telescopes aimed at these slits) are about to register an event of a particle passing through the given slit, from which either the truth or the falsity of the propositions $\left(\{A\}\right)$ and $\left(\{B\}\right)$ can be inferred (such an event qualifies as \textit{a measurement of which-slit information}). Along these lines, the proposition $\left(\{A,B\}\right)$ is logically equivalent to the assertion that the particle has passed through exactly one slit – either $A$ or $B$ (in other words, exactly one outcome has occurred in the measurement).\\

\noindent In standard logic symbols, the proposition $\left(\{A,B\}\right)$ takes the form\smallskip

\begin{equation} \label{2} 
        \left(\{A,B\}\right)
        \equiv
        \Big(
           \big(
              \left(\{A\}\right) \vee \left(\{B\}\right)
           \big)
           \wedge
           \big(
              \neg \left(\{A\}\right) \vee \neg \left(\{B\}\right)
           \big)
        \Big)   
   \;\;\;\;  ,
\end{equation}
\smallskip

\noindent or  it can be represented simpler in Zhegalkin polynomials \cite{Dwinger, Halmos} (defined on the Boolean prototype $\{\bot,\top\} \equiv \{0,1\}$):\smallskip

\begin{equation} \label{3} 
        \left(\{A,B\}\right)
        \equiv
        \left(\{A\}\right) + \left(\{B\}\right) - 2\left(\{A\}\right) \times \left(\{B\}\right)
   \;\;\;\;  ,
\end{equation}
\smallskip

\noindent where the term $\left(\{A\}\right) \times \left(\{B\}\right)$ corresponds to \textit{pairwise interference}.$\,$\footnote{\label{f2}Following Ref. \cite{Barnum}, the definition of interference, which will henceforth be used in the present paper, is not restricted to spatially arranged slits, but is formulated generally for any set of $N$ perfectly distinguishable alternatives.\vspace{5pt}}\\
  
\noindent Let us evaluate the propositional formula (\ref{3}) prior to the measurement. According to the Kochen-Specker theorem \cite{Kochen}, it is impossible to \textit{noncontextually} assign logical values from the two-element set $\{0,1\}$ to $(\{A\})$ and $(\{B\})$ such that in any measurement (of which-slit information) the proposition $(\{A,B\})$ will be true (i.e., there will be only one true outcome in the measurement). Here, by noncontextuality it is implied an assignment of a logical value to a proposition of elementary event in a way that depends only on the proposition itself and not the context (i.e., the logical values of other propositions of elementary events).\\

\noindent E.g., suppose that the \textit{pre-existing} (i.e., pre-measurement) logical values of $(\{A\})$ and $(\{B\})$ are certain and both equal to 1, then in the measurements, in which either slit is blocked, the proposition $(\{A,B\})$ would be true, namely\smallskip

\begin{equation} \label{4} 
        \left(\{A,B\}\right)
        \equiv
        \left\{\!\!
           \begin{array}{l l}
             (\{A\})=1 & \quad \text{$B$ is blocked}\\
             (\{B\})=1 & \quad \text{$A$ is blocked}
           \end{array} \right.
   \;\;\;\;  ,
\end{equation}
\smallskip

\noindent while it would be false in the measurement where the both detectors behind the slits are obtainable.\\

\noindent This means that the pre-measurement proposition $(\{A,B\})$ of exactly one outcome in the two-event space $\{A,B\}$ cannot be written as the sum of the propositions $(\{A\})$ and $(\{B\})$ in the single event spaces $\{A\}$ and $\{B\}$ gotten by making either alternative unobtainable. It is in this sense that second-order quantum interference can be considered irreducible.\\

\noindent One can infer from such an irreducibility that in the two-slit experiment the logical values of the propositions of elementary events $(\{A\})$ and $(\{B\})$ cannot be decided before the measurement, or, in other words, the propositional formula (\ref{3}) cannot a priori be assigned any definite logical value. The proposition $(\{A,B\})$ is taken to be true if there exists evidence (the actual registration of the passage of the particle through the slit) witnessing its truth.$\,$\footnote{\label{f3}This inference is in line with constructive logic where propositional formulae are true due to direct evidence \cite{Bezhanishvili, Bolotin}.\vspace{5pt}}\\

\noindent Let $\Psi(0)=c_A \Psi_A + c_B \Psi_B$ be the pre-measurement wave function of the composite, i.e., particle + detectors, system, where $\Psi_A$ and $\Psi_B$ represent the wave functions from the corresponding single slits, and $\Psi(t)$ be the post-measurement wave function of the system. Along the lines of Gleason's theorem \cite{Gleason, Caves}, assigning \textit{post-measurement} logical values of the propositions $(\{A\})$ and $(\{B\})$ to the wave functions $\Psi_A$ and $\Psi_B$, that is, mapping $\Psi_A$ and $\Psi_B$ to the two-point set $\{0,1\}$, one must choose the following expressions: $(\{A\})\equiv |\langle \Psi_A,\Psi(t) \rangle |^2 = |c_A|^2$ and $(\{B\})\equiv |\langle \Psi_B,\Psi(t) \rangle |^2 = |c_B|^2$, where $|c_A|^2, |c_B|^2 \in \{0,1\}$ and $|c_A|^2+|c_B|^2=1$, $\langle \cdot,\cdot \rangle$ stands for inner product and it is assumed that the set $\{\Psi_A, \Psi_B\}$ is orthonormal.\\

\noindent To demonstrate the association of the pre-measurement propositions $(\{A\})$ and $(\{B\})$ with the prior probabilities $Pr[(\{A\})] \equiv |\langle \Psi_A,\Psi(0) \rangle |^2 = |c_A|^2$ and $Pr[(\{B\})] \equiv |\langle \Psi_B,\Psi(0) \rangle |^2 = |c_B|^2$ where $|c_A|^2, |c_B|^2 \in [0,1]$ (so that $Pr[1]=1$ and $Pr[0]=0$) and $|c_A|^2+|c_B|^2=1$, consider the equal superposition state $c_A=c_B$. Given that prior to the measurement the propositions $(\{A\})$ and $(\{B\})$ are indistinguishable (and for this reason equivalent) in the equal superposition state and after the measurement correspond to the mutually exclusive events, i.e., $(\{A\}) \oplus (\{B\})=1$, one can assign both propositions an equal probability $Pr[(\{A\})]=Pr[(\{B\})]=\frac{1}{2}$ of coming out 1 during the measurement (as long as $(\{A\})$ and $(\{B\})$ cannot be evaluated before the measurement).\\

\noindent In this way, one gets as a necessary consequence of the propositional formula (\ref{3}) that the interference pattern $P_{AB}$ of the two-slit experiment cannot be written as the sum of one-slit patterns $P_A$ and $P_B$ obtained by blocking each of the slits $A$ and $B$, namely,\smallskip

\begin{equation} \label{5} 
        P_{AB} - P_A - P_B \neq 0
   \;\;\;\;  .
\end{equation}
\smallskip

\subsection{A triple slit experiment}\label{Three-slit}

\noindent Now, let us consider a triple slit experiment. In contrast to the preceding subsection \ref{Two-slit}, the complication here is that in general the operation of exclusive disjunction is defined to be true if \textit{an odd number of its arguments are true} and false otherwise \cite{Simpson}. So, the chain of two operations $\oplus$, namely $(\{A\})\oplus(\{B\})\oplus(\{C\})$, will be true if exactly one outcome is true as well as all three outcomes are simultaneously true:\smallskip

\begin{equation} \label{6} 
        a \oplus b \oplus c
        \equiv
        a + b + c - 2\left(a b+a c+b c\right) + 4 a b c =
        \left\{\!\!
           \begin{array}{l l}
            1 & \quad \text{$a = b = c = 1$}\\
            1 & \quad \text{one of $a,b,c$ is $1$}\\
            0 & \quad \text{else}
        \end{array} \right.
   \;\;\;\;  ,
\end{equation}
\smallskip

\noindent where, for brevity, the following notations are used $a\! \equiv \! (\{A\})$, $b \! \equiv \! (\{B\})$ and $c \! \equiv \! (\{C\})$. The chain $a \oplus b \oplus c$ can also be presented as the following sum\smallskip

\begin{equation} \label{7} 
        a \oplus b \oplus c
        \equiv
        \Upsilon_{\!ABC} + 4 a b c
   \;\;\;\;  ,
\end{equation}
\smallskip

\noindent in which the first term is\smallskip

\begin{equation} \label{8} 
       \Upsilon_{\!ABC}
        \equiv
        \left(\{A,B\}\right) + \left(\{B,C\}\right) + \left(\{A,C\}\right) - a - b - c
   \;\;\;\;  ,
\end{equation}
\smallskip

\noindent where $\left(\{A,B\}\right)$, $\left(\{B,C\}\right)$ and $\left(\{A,C\}\right)$ are the propositions of exactly one outcome in the two-event spaces $\{A,B\}$, $\{B,C\}$ and $\{A,C\}$, at the same time as the term $abc$ corresponds to \textit{third-order interference}.\\

\noindent The truth table of the chain $a \oplus b \oplus c$  implies that the proposition $(\{A,B,C\})$ of exactly one outcome in the triple event space $\{A,B,C\}$ should take the form\smallskip

\begin{equation} \label{9} 
        \left( \{A,B,C\} \right)
        \equiv
        \left( a \oplus b \oplus c \right) \left( 1 \oplus a b c \right)
         =
        \Upsilon_{\!ABC} \left( 1 \oplus a b c \right)
         =
        \left\{\!\!
           \begin{array}{l l}
            1 & \quad \text{one of $a,b,c$ is $1$}\\
            0 & \quad \text{else}
        \end{array} \right.
   \;\;\;\;  ,
\end{equation}
\smallskip

\noindent where $1$ stand for the purely true term.\\

\noindent In consonance with the case of the two-slit experiment, by blocking one of the three slits – say $A$ – one infers that the proposition $a$ is false and thus $a b c = 0$; but even so, the logical values of two other propositions – $b$ and $c$ – remain indeterminate (as they cannot be a priori evaluated, and blocking one out of the three slits cannot qualify as a measurement of which-slit information). Hence, by making any one of the alternatives $A$, $B$ and $C$ unobtainable, the pre-measurement proposition of exactly one outcome in the 3-event space can be rewritten in terms of the pre-measurement propositions of exactly one outcome in the two-event spaces, namely, $\left( \{A,B,C\} \right) = \Upsilon_{\!ABC}$.\\

\noindent This implies that the interference pattern $P_{ABC}$ obtained in the three-slit experiment can be expressed in terms of the two-slit patterns obtained by blocking one of the slits $A$, $B$, $C$:\smallskip

\begin{equation} \label{10} 
        P_{ABC} - P_{AB} - P_{BC} - P_{AC} + P_A + P_B + P_C = 0
   \;\;\;\;  .
\end{equation}
\smallskip

\noindent If in every measurement of which-slit information in the three-slit experiment exactly one outcome always occurs, then after the measurement one would get $\left( \{A,B,C\} \right) = a + b + c = 1$ and, accordingly, in place of (\ref{10}) one would find\smallskip

\begin{equation} \label{11} 
        P_{ABC} - P_A - P_B - P_C = 0
   \;\;\;\;  .
\end{equation}
\smallskip

\subsection{Four- and five-slit experiments}\label{More-slit}

\noindent In experiments involving a 4-path interferometer, the chain $\mathrm{XOR}_4$ of operations of exclusive disjunction $\oplus$ is true if exactly one outcome is true or any three outcomes (out of 4) are simultaneously true:\smallskip

\begin{equation} \label{12} 
        \mathrm{XOR}_4 
        \equiv
        a \oplus b \oplus c \oplus d
        =
        \left\{\!\!
           \begin{array}{l l}
            1 & \quad \text{any 3 of $a,b,c,d$ are $1$}\\
            1 & \quad \text{one of $a,b,c,d$ is $1$}\\
            0 & \quad \text{else}
        \end{array} \right.
   \;\;\;\;  ,
\end{equation}
\smallskip

\noindent where in addition to the previous notations the new one is used $d \equiv (\{D\})$. With the ordinary operations of arithmetic, the chain $\mathrm{XOR}_4$ may be expressed as\smallskip

\begin{equation} \label{13} 
        \mathrm{XOR}_4 
        \equiv
        \Upsilon_{\!ABCD} + 4 \left( abc + abd + acd + bcd \right) - 8 abcd
   \;\;\;\;  .
\end{equation}
\smallskip

\noindent In this expression, the term $abcd$ corresponds to four-order interference and $(\cdot)$ to third-order interference, $\Upsilon_{\!ABCD}$ represents the propositional formula\smallskip

\begin{equation} \label{14} 
        \begin{array}{c}
            \Upsilon_{\!ABCD} \equiv \left(\{A,B\}\right) + \left(\{A,C\}\right) + \left(\{A,D\}\right) +\left(\{B,C\}\right) + \left(\{B,D\}\right) + \left(\{C,D\}\right) \\
            - 2\left( a + b + c + d \right)
        \end{array}
   \;\;\;\;  ,
\end{equation}
\smallskip

\noindent in which $(\{\cdot, \cdot\})$ denote propositions of exactly one outcome in two-event spaces $\{\cdot, \cdot\}$.\\

\noindent In relation to the truth table of the chain $\mathrm{XOR}_4$, the proposition $(\{A,B,C,D\})$ of exactly one outcome in the quadripartite event space $\{A,B,C,D\}$ must take the following form\smallskip

\begin{equation} \label{15} 
        \left(\{A,B,C,D\}\right)
        \equiv
        \mathrm{XOR}_4 (1 \oplus abc) (1 \oplus abd) (1 \oplus acd) (1 \oplus bcd)
   \;\;\;\;  .
\end{equation}
\smallskip

\noindent Again, by making any two of the alternatives $A$, $B$ , $C$ and $D$ unobtainable, for the pre-measurement proposition $\left(\{A,B,C,D\}\right)$ one gets $\left(\{A,B,C,D\}\right) \equiv \Upsilon_{\!ABCD}$, where the term $\Upsilon_{\!ABCD}$ cannot a priori be assigned any definite logical value.\\

\noindent This indicates that the probability $P_{ABCD}$ of the click of a detector localized in a certain area of the screen behind the 4-slit setup can be computed from contributions of pairs of slits only, that is,\smallskip

\begin{equation} \label{16} 
        P_{ABCD} - P_{AB} - P_{AC} - P_{AD} - P_{BC} - P_{BD} - P_{CD} + 2P_A + 2P_B + 2P_C + 2P_D = 0
   \;\;\;\;  .
\end{equation}
\smallskip

\noindent Provided that in every measurement of which-slit information in the 4-slit experiment exactly one outcome always occurs, after the measurement the expression (\ref{16}) will result in\smallskip

\begin{equation} \label{17} 
        P_{ABCD} - \left(P_A + P_B + P_C + P_D\right) = 0
   \;\;\;\;  .
\end{equation}
\smallskip

\noindent Analogously, the proposition of exactly one outcome in the 5-event space $\{A,B,C,D,E\}$ must take the form\smallskip

\begin{equation} \label{18} 
        \left(\{A,B,C,D,E\}\right)
        \equiv
        \left(a \oplus b \oplus c \oplus d \oplus e\right) \Delta_{ABCDE}
   \;\;\;\;  ,
\end{equation}
\smallskip

\noindent in which $e \equiv (\{E\})$ and the factor $\Delta_{ABCDE}$ denotes\smallskip

\begin{equation} \label{19} 
        \begin{array}{c}
            \Delta_{ABCDE} \equiv (1\oplus abc)(1\oplus abd)(1\oplus abe)(1\oplus acd)(1\oplus ace) \\
            (1\oplus ade)(1\oplus bcd)(1\oplus bce)(1\oplus bde)(1\oplus cde)
        \end{array}
   \;\;\;\;  .
\end{equation}
\smallskip

\noindent Presenting the chain $\mathrm{XOR}_5 \equiv a \oplus b \oplus c \oplus d \oplus e$ in the form of ordinary arithmetical operations\smallskip

\begin{equation} \label{20} 
        \begin{array}{c}
            \mathrm{XOR}_5 \equiv \Upsilon_{\!ABCDE}\\
            + 4 \left( abc+abd+abe+acd+ace+ade+bcd+bce+bde+cde \right) \\
            -  8 \left( abcd+abce+abde+acde+bcde \right) + 16 abcde
        \end{array}         
   \;\;\;\;  ,
\end{equation}
\smallskip

\noindent where the term $\Upsilon_{\!ABCDE}$ is\smallskip

\begin{equation} \label{21} 
        \begin{array}{c}
            \Upsilon_{\!ABCDE} \equiv \left(\{A,B\}\right) + \left(\{A,C\}\right) + \left(\{A,D\}\right) + \left(\{A,E\}\right) \\
            +\left(\{B,C\}\right) + \left(\{B,D\}\right) + \left(\{B,E\}\right) + \left(\{C,D\}\right) + \left(\{C,E\}\right) + \left(\{D,E\}\right) \\
            - 3 \left( a + b + c + d + e \right)
        \end{array}
   \;\;\;\;  ,
\end{equation}
\smallskip

\noindent and given that $(x^2)=(x)$ and $(x)\oplus (x)=0$, the proposition $(\{A,B,C,D,E\})$ can be rewritten as\smallskip

\begin{equation} \label{22} 
        \left(\{A,B,C,D,E\}\right)
        \equiv
        \mathrm{XOR}_5 \Delta_{ABCDE}
        =
       \Upsilon_{\!ABCDE} \Delta_{ABCDE}
   \;\;\;\;  .
\end{equation}
\smallskip

\noindent Once again, excluding any three alternatives out of five ones, one gets $\Delta_{ABCDE}=1$ and so $\left(\{A,B,C,D,E\}\right) \equiv \Upsilon_{\!ABCDE}$, which suggests that the interference pattern $P_{ABCDE}$ in the experiment with a 5-path interferometer can be explained using the second order interference terms only, namely,\smallskip

\begin{equation} \label{23} 
        \begin{array}{c}
            P_{ABCDE} - P_{AB} - P_{AC} - P_{AD} - P_{AE} - P_{BC} - P_{BD} - P_{BE} - P_{CD} - P_{CE} - P_{DE} \\
            - 3 \left(P_A+P_B+P_C+P_D+P_E \right) = 0
        \end{array}
   \;\;\;\;  .
\end{equation}
\smallskip

\section{Discussion}\label{Discussion}

\noindent As it follows from the previous section, the chain $\mathrm{XOR}_N$ of operations of exclusive disjunction of an arbitrary length $N>2$, namely, $\mathrm{XOR}_N \equiv x_1 \oplus \cdots \oplus x_i \oplus \cdots \oplus x_N$, where $x_i \equiv (\{X_i\})$ denote the propositions of elementary events $\{X_i\} \subset \{X_1,\cdots,X_i,\cdots,X_N\}$, in addition to the term $\Upsilon_{\!N}$, which contains the propositions $(\{X_i,X_j\}) \equiv x_i \oplus x_j$ of exactly one outcome in the two-event spaces $\{X_i,X_j\}$, i.e.,\smallskip

\begin{equation} \label{24} 
        \Upsilon_{\!N}
        \equiv
        \sum_{i<j}^{N} \left(\{X_i,X_j\}\right) - \left( N-2 \right) \sum_{i}^{N} x_i
   \;\;\;\;  ,
\end{equation}
\smallskip

\noindent has also the terms $I_{3}(x_i x_j x_k)$, $I_{4}(x_i x_j x_k x_l)$, $I_{5}(x_i x_j x_k x_l x_m)$, and the like, corresponding to third- and higher-order interference:\smallskip

\begin{equation} \label{25} 
        \mathrm{XOR}_N
        \equiv
        \Upsilon_{\!N} + I_{3}\left(x_i x_j x_k\right) + I_{4}\left(x_i x_j x_k x_l\right) + I_{5}\left(x_i x_j x_k x_l x_m\right) + \dots
   \;\;\;\;  .
\end{equation}
\smallskip

\noindent On the other hand, for the proposition $(\{X_1,\dots,X_N\})$ to be logically equivalent to the assertion that the particle has passed through exactly one slit $X_i$ in the $N$-slit experiment, that is, exactly one outcome $\{X_i\}$ has occurred in the $N$-event space $\{X_1,\dots,X_N\}$, besides the chain $\mathrm{XOR}_N$ this proposition must also include the factor $\Delta_N$\smallskip

\begin{equation} \label{26} 
        \Delta_N
        \equiv
        \prod_{i<j<k}^{N}\!\! \left( 1\oplus x_i x_j x_k \right)
   \;\;\;\;  ,
\end{equation}
\smallskip

\noindent which ensures that the given proposition will be true when exactly one $x_i$ will be recorded as true while others $x_{j \neq i}$ will not, that is:\smallskip

\begin{equation} \label{27} 
        \left(\{X_1,\dots,X_N\}\right)
        \equiv
        \mathrm{XOR}_N \Delta_N
        =
        \left\{\!\!
           \begin{array}{l l}
            1 & \quad \text{$x_i=1$, $x_{j \neq i}=0$}\\
            0 & \quad \text{else}
        \end{array} \right.
   \;\;\;\;  .
\end{equation}
\smallskip

\noindent Yet, when multiplied by the factor $\Delta_N$ all higher-order interference terms $I_{3}(x_i x_j x_k)$, $I_{4}(x_i x_j x_k x_l)$, $I_{5}(x_i x_j x_k x_l x_m)$, … in the chain $\mathrm{XOR}_N$ vanish, explicitly,\smallskip

\begin{equation} \label{28} 
        \mathrm{XOR}_N \Delta_N
        =
        \Upsilon_{\!N} \Delta_N
   \;\;\;\;  .
\end{equation}
\smallskip

\noindent As a result, by making some alternatives unobtainable, i.e., by making the factor $\Delta_N$ equal to 1, the pre-measurement proposition $(\{X_1,\dots,X_N\})$ of exactly one outcome in the $N$-event space can be written in terms of the pre-measurement propositions $(\{X_i,X_j\})$ of exactly one outcome in the double-event spaces $\{X_i,X_j\}$ corrected for over-counting by subtracting suitable multiples of the propositions $x_i$ of elementary events $\{X_i\}$:\smallskip

\begin{equation} \label{29} 
        \left(\{X_1,\dots,X_N\}\right)
        =
        \sum_{i<j}^{N} \left(\{X_i,X_j\}\right) - \left( N-2 \right) \sum_{i}^{N} x_i
   \;\;\;\;  .
\end{equation}
\smallskip

\noindent This implies that the interference pattern $P_{1 \dots N}$ in the $N$-slit set-up can be represented in terms of the sum of the double-slit patterns, namely,\smallskip

\begin{equation} \label{30} 
        P_{1 \dots N}
        -
        \sum_{i<j}^{N} P_{ij} + \left( N-2 \right) \sum_{i}^{N} P_i
        = 0
   \;\;\;\;  .
\end{equation}
\smallskip

\noindent In a nutshell: The reason for quantum theory to only predict second-order interference is the assumption (underlying the Born rule) that in each measurement in the $N$-event space exactly one outcome may only occur. Thus, if the pattern $P_{1 \dots N}$ were to fail to follow the equality (\ref{30}), then such a fundamental assumption would turn out to be wrong.\\

\noindent Otherwise stated, the existence of third- or higher-order interference would contradict the proposition that \textit{measurements always yield definite results}. This means that ongoing experiments, which look for possible higher-order interference, in fact are testing the truthfulness of exactly that proposition.\\

\bibliographystyle{References}
\bibliography{References}

\end{document}